\documentclass[aip,amsmath,amssymb,preprint]{revtex4-1}

\usepackage{graphicx,amssymb,bm,amsfonts,amsmath,graphics,color,epstopdf}
\usepackage[bookmarks=false,pdfstartview=FitH,colorlinks=true,citecolor=blue,linkcolor=blue]{hyperref}
\usepackage{lineno}
\usepackage{physics}
\begin{document}

\title{Competition between magnetic order and charge localization in Na$_{2}$IrO$_{3}$ thin crystal devices}

\author{Josue Rodriguez}
\affiliation{Department of Physics and Astronomy, California State University Long Beach, Long Beach, California 90840, USA}

\author{Gilbert Lopez}
\affiliation{Department of Physics, 
University of California Berkeley, California 94720, USA}

\author{Samantha Crouch}
\affiliation{Department of Physics and Astronomy, California State University Long Beach, Long Beach, California 90840, USA}

\author{Nicholas P. Breznay}
\affiliation{Department of Physics, 
Harvey Mudd College, Claremont, California 91711, USA}

\author{Robert Kealhofer}
\affiliation{Department of Physics, 
University of California Berkeley, California 94720, USA}

\author{Vikram Nagarajan}
\affiliation{Department of Physics, 
University of California Berkeley, California 94720, USA}

\author{Drew Latzke}
\affiliation{Department of Physics, 
University of California Berkeley, California 94720, USA}

\author{Francisco Ramirez}
\affiliation{Department of Physics and Astronomy, California State University Long Beach, Long Beach, California 90840, USA}

\author{Naomy Marrufo}
\affiliation{Department of Physics and Astronomy, California State University Long Beach, Long Beach, California 90840, USA}

\author{Peter Santiago}
\affiliation{Department of Chemistry and Biochemistry, California State University Long Beach, Long Beach, California 90840, USA}

\author{Jared Lara}
\affiliation{Department of Physics and Astronomy, California State University Long Beach, Long Beach, California 90840, USA}

\author{Amirari Diego}
\affiliation{Department of Physics and Astronomy, California State University Long Beach, Long Beach, California 90840, USA}

\author{Everardo Molina}
\affiliation{Department of Physics and Astronomy, California State University Long Beach, Long Beach, California 90840, USA}

\author{David Rosser}
\affiliation{Department of Physics and Astronomy, California State University Long Beach, Long Beach, California 90840, USA}

\author{Hadi Tavassol}
\affiliation{Department of Chemistry and Biochemistry, California State University Long Beach, Long Beach, California 90840, USA}

\author{Alessandra Lanzara}
\affiliation{Department of Physics, 
University of California Berkeley, California 94720, USA}

\author{James G. Analytis}
\affiliation{Department of Physics, 
University of California Berkeley, California 94720, USA}

\author{Claudia Ojeda-Aristizabal}
\affiliation{Department of Physics and Astronomy, California State University Long Beach, Long Beach, California 90840, USA}

\date{\today}

\begin{abstract}
Spin orbit assisted Mott insulators such as sodium iridate (Na$_2$IrO$_3$) have been an important subject of study in the recent years. In these materials, the interplay of electronic correlations, spin-orbit coupling, crystal field effects and a honeycomb arrangement of ions bring exciting ground states, predicted in the frame of the Kitaev model. The insulating character of Na$_2$IrO$_3$ has hampered its integration to an electronic device, desirable for applications, such as the manipulation of quasiparticles interesting for topological quantum computing. Here we show through electronic transport measurements supported by Angle Resolved Photoemission Spectroscopy (ARPES) experiments, that electronic transport in Na$_2$IrO$_3$ is ruled by variable range hopping and it is strongly dependent on the magnetic ordering transition known for bulk Na$_2$IrO$_3$, as well as on external electric fields. Electronic transport measurements allow us to deduce a value for the localization length and the density of states in our Na$_2$IrO$_3$ thin crystals devices, offering an alternative approach to study insulating layered materials.       
\end{abstract}


\maketitle

%

Sodium iridate (Na$_2$IrO$_3$) and other 5d iridates have been widely studied in recent years as they have been identified as experimental realizations of the Kitaev model, an exactly solvable model that describes a set of spin-1/2 moments in a honeycomb lattice with highly anisotropic exchange interaction and exciting ground states such as spin liquids \cite{Chun, Singh, Winter}. In real materials such as Na$_2$IrO$_3$, the existence of Heisenberg and off diagonal interactions benefit the appearance of magnetically ordered ground states over a spin liquid. Recent studies have explored the effect of external pressure and high magnetic fields on the ground states of Na$_2$IrO$_3$ \cite{6Proposal,7Proposal,8Proposal}. First principle calculations \cite{6Proposal} have found that under high pressure, Na$_2$IrO$_3$ goes through successive structural and magnetic phase transitions, some of them zigzag magnetic ordered and some nonmagnetic, all with low energy excitations that can be well described by j$_{eff}$=1/2 states and in particular having some phases (with a structure corresponding to space group $P\Bar{1}$)\cite{6Proposal} resembling a gapped spin liquid Kitaev state. Similarly,  magnetic torque measurements have found that at magnetic fields up to 60T, long range spin correlations functions decay rapidly pointing to a field-induced quantum spin liquid \cite{8Proposal}. In the absence of high pressure or magnetic fields, recent inelastic x-ray scattering measurements have reported a proximate spin liquid regime above the long-range ordering temperature, where fractional excitations are emergent, revealed by spin-spin correlations restricted to nearest neighbor sites, among other factors \cite{Ravelli}.

Older and newer studies on Na$_2$IrO$_3$ have been limited to probes that require bulk crystals, where the insulating character of Na$_2$IrO$_3$ has hampered its integration to an electronic device, advantageous for the study and manipulation of quasiparticle excitations, of interest in topological quantum computing\cite{Kitaev,Kasahara}. Here we report electronic transport measurements on thin crystals of Na$_2$IrO$_3$, where electronic transport is ruled by variable range hopping at temperatures above the magnetic ordering transition known for bulk Na$_2$IrO$_3$ ($\sim 15$~K) and electric field assisted above a critical applied field. Fits of our data to these two mechanisms allow us to deduce the localization length as well as the density of states at the Fermi level in our Na$_2$IrO$_3$ devices. Our Angle Resolved Photoemission Spectroscopy (ARPES) measurements show a non-vanisihing density of states at the Fermi level, consistent with our electronic transport experiments.

Bulk crystals of Na$_2$IrO$_3$ were grown mixing elemental Ir ($99.9\%$ purity, BASF) with Na$_2$CO$_3$ ($99.9999\%$ purity, Alfa-Aesar) in a $1:1.05$ molar ratio. The mixture was ground for several minutes and pressed into a pellet at approximately $3,000$ psi. The pellet was subsequently warmed in a furnace to $1050\,^{\circ}\mathrm{C}$ and held at this temperature for 48 hours, before being cooled to $900,^{\circ}\mathrm{C}$ over 24 hours and finally furnace-cooled. Single crystals of more than one square millimeter were collected from the surface of the pellet. Crystals were exfoliated onto a Si/SiO$_{2}$(280~nm) substrate and immediately coated with e-beam resist, minimizing the exposure to air. 

\begin{figure}
\includegraphics[width=\linewidth]{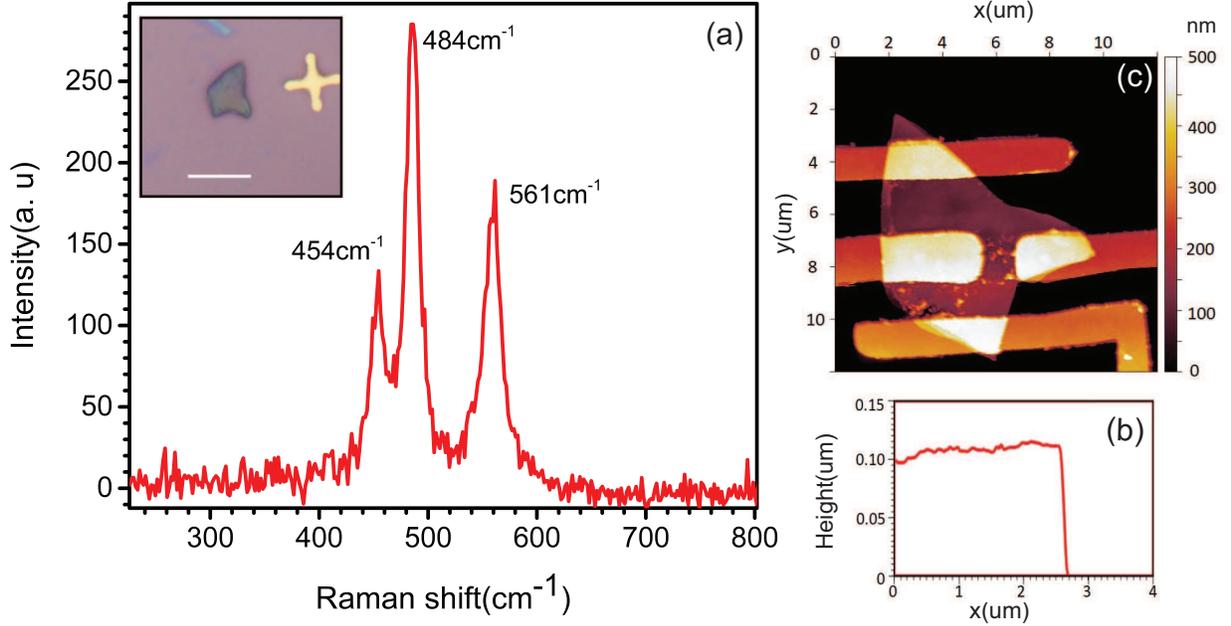}
\caption{\label{fig:Fig1} \textbf{(a)} Raman spectra of an exfoliated thin crystal of Na$_{2}$IrO$_{3}$. Inset: Optical image of the crystal used for device fabrication (scale bar 10~$\mu$m). \textbf{(b)} Atomic force microscopy scan of the electronic device. Electrodes used for the two probe measurements were middle left and bottom. \textbf{(c)} AFM height profile of the device, crystal is approximately 100~nm thick.}
\end{figure}

Some wafers with exfoliated crystals were left uncoated and promptly analyzed through Raman spectroscopy. Raman scans were made over a range of 56~$cm^{-1}$ to 800~$cm^{-1}$. Three Raman active modes were observed at 454~$cm^{-1}$, 484~$cm^{-1}$ and 561~$cm^{-1}$ (Figure~\ref{fig:Fig1}a) consistent with the reported first-order Raman modes of bulk sodium iridate \cite{Gupta}. Exfoliated crystals were about $\sim$100~nm thick, measured through atomic force microscopy (Figure ~\ref{fig:Fig1}b and c). Electrodes were patterned on the ebeam resist-coated wafers using standard electron beam lithography, previous to the deposition of Ti(4~nm)/Au(140~nm) using electron beam evaporation. 

\begin{figure}
\includegraphics[width=\linewidth]{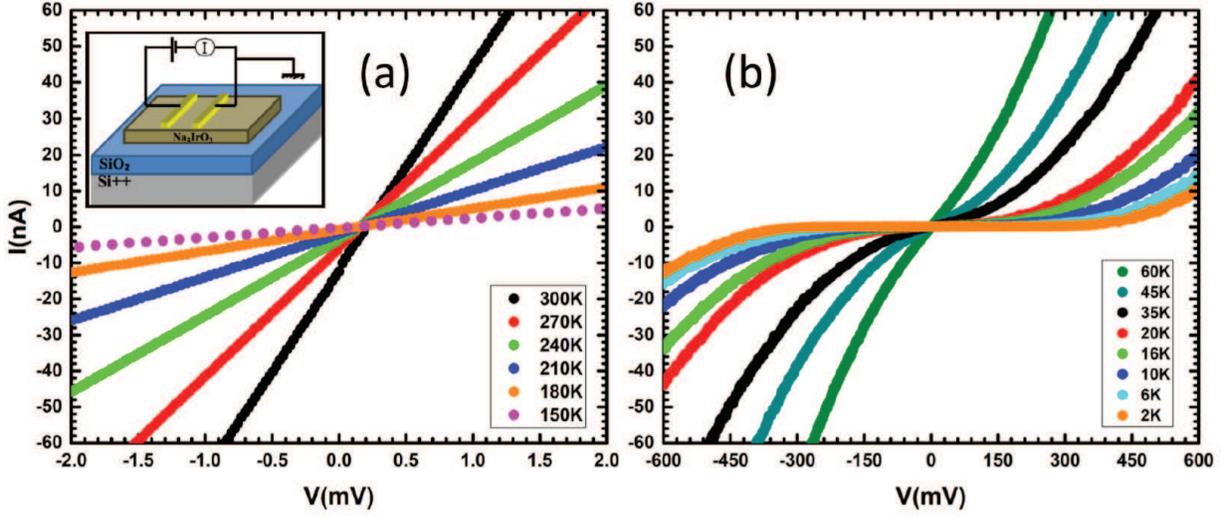}
\caption{\label{fig:IV}\textbf{(a)} IV characteristics of the Na$_2$IrO$_3$ device at high temperatures. Inset: schematics of the electronic device. \textbf{(b)} Non-linear IV curves at low temperatures down to 2~K.}
\end{figure}

Current-voltage (IV) characteristics were measured (two probe) in a closed cycle cryostat (Oxford Instruments) from 300~K down to 1.5~K. As shown in Figure~\ref{fig:IV}a, at high temperatures the IV curves show an ohmic-like behavior. As the temperature is lowered, the slope of the IV curves decreases indicating an increase of the device's resistance. At sufficient low temperatures, the IV characteristics become non-linear (Figure \ref{fig:IV}b). This behavior is consistent with the known Mott-insulating character of Na$_2$IrO$_3$. Previous numerical works that consider both Coulomb interactions and spin-orbit coupling \cite{Comin} have been able to explain the observed insulating gap in Na$_2$IrO$_3$\cite{Comin,Sohn} considering a Coulomb repulsion U (3eV) that is larger than the effective bandwidth ($\approx$ 1eV), supporting the idea of Na$_2$IrO$_3$ being a spin-orbit assisted Mott insulator.

\begin{figure}
\hspace{-20mm}
\includegraphics[width=100mm]{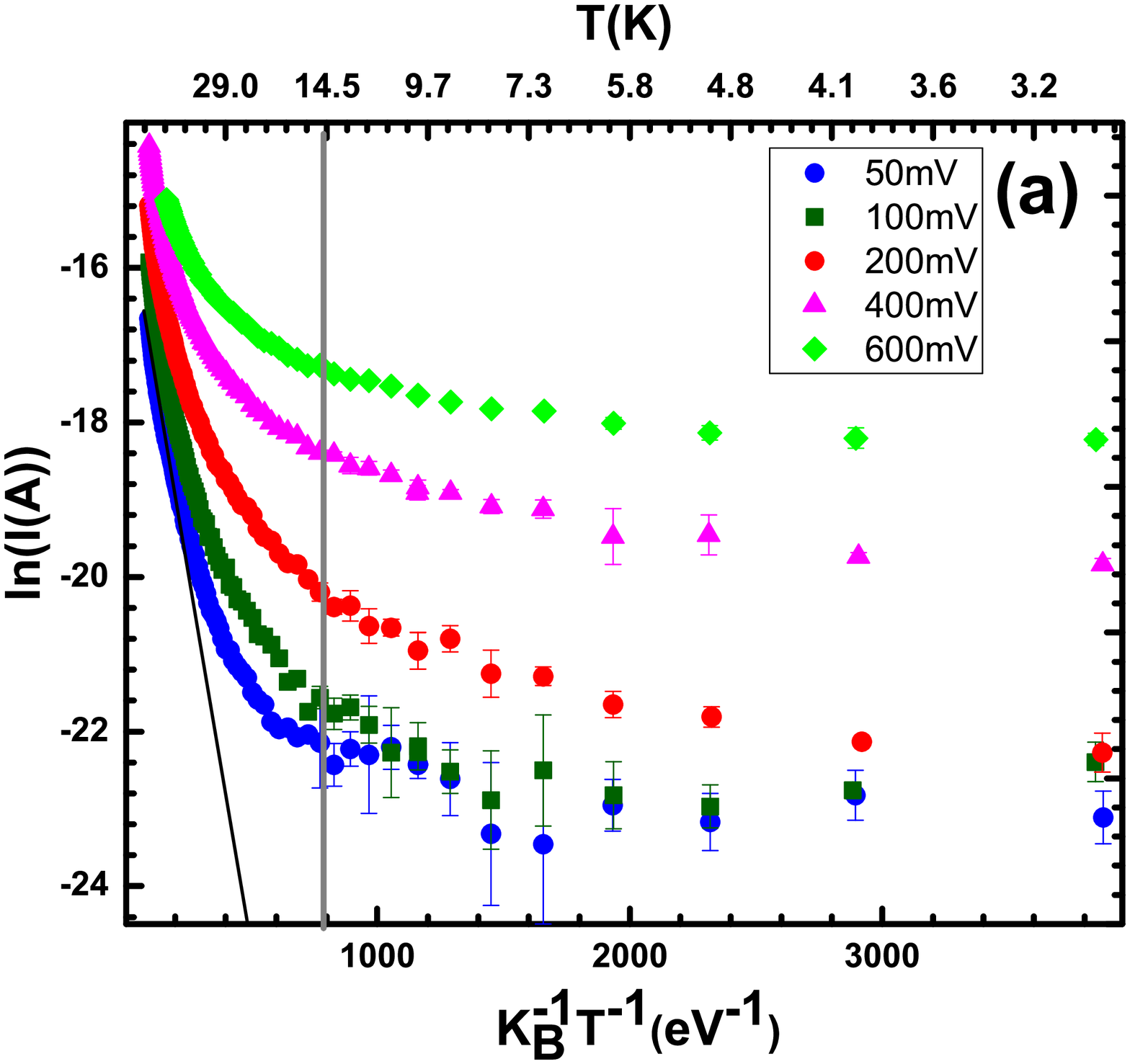} 
\hspace{-20mm}
\includegraphics[width=100mm]{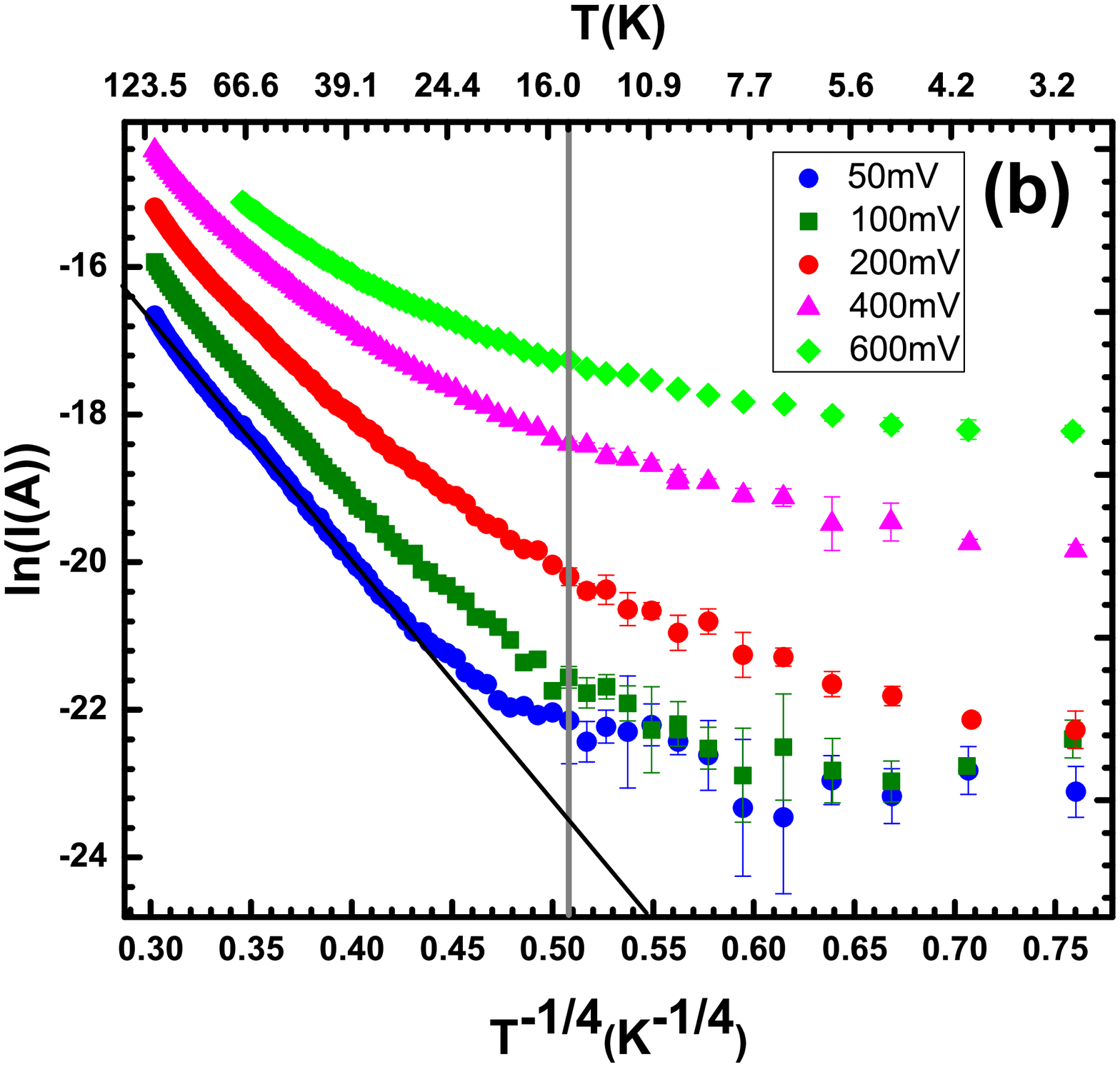}
\hspace{-20mm}
\caption{Temperature dependence of the current $I$ at different bias voltages 50~mV, 200~mV, 400~mV and 600~mV, extracted from the measured IV characteristics. Temperature range is 3K - 120K\textbf{(a)} $I$ versus $1/k_{B}T$ and \textbf{(b)} $I$ versus $T^{-1/4}$ on a semi-log plot.}
    \label{Ther-VRH}
\end{figure}

Being an accurate measurement of the conductance of the sample at zero bias non-accessible at low temperatures, we tracked the change of the current across the sample at non-zero bias as a function of the temperature in a range of 3K-120K (see Fig. \ref{Ther-VRH}), and compared it to the thermal activation theory, characterized by an Arrhenius law,
\begin{equation} \label{eq:TA}
    I(T) \propto I_o e^{-E_{A}/ k_{B}T}
\end{equation}
where E$_{A}$ is the activation energy and $k_{B}$ is Boltzmann constant. Figure~\ref{Ther-VRH}a shows $I$ versus $1/k_{B}T$ on a semi-log plot for different bias voltages. While the data fits in small range of temperatures to a thermal activated behavior (at around 120K), it follows in a more extended range a three-dimensional variable range hopping (VRH) mechanism \cite{Mott} ($\approx15K$-120K), as shown in Figure \ref{Ther-VRH}b. Mott's variable range hopping (VRH) law considers 3-dimensional hopping between remote sites whose energy levels happen to be close to the Fermi level $\mu$, as spatial neighboring sites have larger resistances \cite{BookES9}.

Typically, the hopping length increases with lowering temperature, therefore the name of variable range hopping \cite{BookES9}. In general, Mott's VRH is described by 

\begin{equation} \label{eq:VRH}
    I(T) \propto I_oe^{-(T_{o}/T)^{\nu}},
\end{equation}
with $\nu=1/4$ for three-dimensional VRH and $T_o$ the energy scale related to localization length of the charge carriers,  
\begin{equation} \label{eq:To}
T_o=21.2/k_Ba^3g_0    
\end{equation}
with $a$ the localization length and $g_0$ the density of states at the Fermi level. A fit of the data at low bias (50mV) to three-dimensional VRH yields $T_o^{1/4}=31K^{1/4}$ (see Fig. \ref{Ther-VRH}b), similar to the reported values for heteroepitaxial Na$_2$IrO$_3$ thin films \cite{Jenderka}.

Inspecting more carefully the data in Fig. \ref{Ther-VRH}b at different bias voltages, two main effects are visible. First, there is a change in the trend of the curves near $\sim15K$ (grey line in Figures~\ref{Ther-VRH}a and b), corresponding to the critical temperature of the long range antiferromagnetically ordered state reported for bulk crystals after magnetic susceptibility and heat capacity measurements \cite{Singh2010, Liu2011}, neutron and x-ray diffraction\cite{Ye}. The onset of long range magnetic ordering seems here to have an effect on the charge transport mechanism in the thin crystals of Na$_2$IrO$_3$, possibly suppressing VRH below the transition temperature. A similar phenomenon has been observed in the temperature dependence of the resistivity of heteroepitaxially grown thin films of Na$_{2}$IrO$_{3}$ \cite{Jenderka}. Also, non-equilibrium optical measurements have shown signature of a long range ordered state in photoinduced reflectance measurements \cite{Hinton}. Second, as the bias voltage increases, $\ln(I)$ vs $T^{-1/4}$ deviates from a linear dependence, indicating also a divergence from a VRH mechanism.

We can understand effect 1 by taking a closer look to VRH. Following the model by Miller and Abrahams \cite{BookES4}, the resistivity $R_{ij}$ resulting from hopping over sites ij is given by:
\begin{equation}\label{MAeqn}
    R_{ij}=R_{ij}^0\exp\Big(\frac{2r_{ij}}{a}+\frac{\epsilon_{ij}}{k_BT}\Big), 
\end{equation}
where $r_{ij}$ is the spatial distance to the nearest neighboring site and $\epsilon_{ij}$ correspond to the difference of the hopping sites energies $\epsilon_i$ and $\epsilon_j$ lying in a narrow band near the Fermi energy, whose width $\epsilon_o$ decreases with temperature and depends on the density of states at the Fermi level g$_o$ and localization length $a$, 
\begin{equation}\label{epsilon}
    \epsilon_o=\frac{\big(k_BT\big)^{3/4}}{\big(g_oa^3\big)^{1/4}}
\end{equation}
Hopping between sites depends on both the spatial and energetic separation of the sites. If $2r_{o}/a>>1$ (with $r_o$ the average spatial distance to the nearest neighbor empty size), the conduction mechanism is reduced to nearest neighbor hopping. If on the other hand $2r_o/a$ is of the order or less than unity, the second term in equation \ref{MAeqn} contributes the most to the resistance and hops to sites that are further away in space but closer in energy are favored. Hopping between sites is in this case variable range, which corresponds to the transport mechanism in our thin crystals of Na$_2$IrO$_3$ above the magnetic ordering transition. 
In fact, Mott's VRH relation (equation \ref{eq:VRH}) is derived by finding the energy $\epsilon_{ij}$ from equation \ref{MAeqn} at which the resistivity is minimum. In the case of a magnetic ordered state, previous works have introduced an extension of Mott's VRH model in which the hopping energy $\epsilon_{ij}$ has an additional term related to the relative magnetization of the hopping sites \cite{Wagner}. It is however unclear if associating a cost on the hopping energy based on the spin orientation of the hopping sites is valid, as carriers have tendency to align their spin to the localized spin \cite{CommentWagner, DeGennes}. Based on our data, we believe that electronic transport in our Na$_2$IrO$_3$ devices in the magnetic ordered regime goes beyond Mott's VRH model.

Effect 2 caused by the bias voltage, can be unveiled by adding the effect of an electric field $E$ to equation \ref{MAeqn}, obtaining
\begin{equation}\label{MAeqn2}
   R_{ij}=R_{ij}^0\exp\Bigg(\frac{2r_{ij}}{a}+\frac{\epsilon_{ij}-er_{ij}E}{k_BT}\Bigg),
\end{equation}

\begin{figure}
\includegraphics[width=10cm]{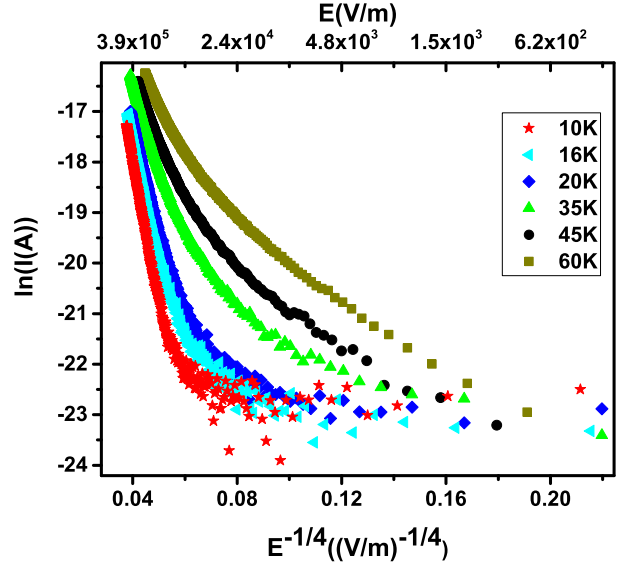}
\caption{\label{Field} Electric field dependence of the current I versus E$^{-1/4}$ in a temperature range 10K-60K} 
\end{figure}

At high electric field strengths, where $|er_{ij}E| \geq \epsilon_{ij}$, hopping in the direction of the electric field will compensate the difference energy between hopping sites, $\epsilon_{ij}$. Therefore hops between sites that are closer in energy are no longer energetically favorable and the first term in  equation \ref{MAeqn2} dominates. The average distance between hopping sites $r$ is expected to satisfy $|erE| = \epsilon_{ij}$. Taking  $\epsilon_{ij} \sim 1/r^{3}g_{o}$ (where $g_{o}$ is the density of states at the Fermi level) the average distance between hopping sites becomes $r \sim (1/eEg_{o})^{1/4}$ giving rise to a field assisted transport mechanism \cite{Shklovskii}
\begin{equation}\label{MAeqn3}
   R_{ij}=R_{ij}^0\exp(\bigg(\frac{E_{o}}{E}\bigg) ^{1/4}) \qquad\text{or}\qquad
   I\propto I_{o}\exp(-\bigg(\frac{E_{o}}{E}\bigg) ^{1/4}),
\end{equation}
with $E_o$ the characteristic localization field.
\begin{equation}\label{Eo}
E_{o} = 1/e g_{o}a^{4}
\end{equation}
Above a critical external electric field, carrier hopping is therefore mediated by the electric field and there is no longer a clear VRH mechanism, as observed for the highest fields in Figure \ref{Ther-VRH}b. In fact, there is an electric field assisted transport at high fields that can be observed in the data, which is accentuated at lower temperatures as is shown in Figure \ref{Field}, where IV curves at different temperatures are represented following relation \ref{MAeqn3}. Below a critical electric field, transport is no longer field assisted. A fit of the data represented in Fig.\ref{Field} to equation \ref{MAeqn3} allow us to extract $E_o$. In particular, for 20K we obtain $E_o^{1/4}=190(V/m)^{1/4}$.

From the definitions of $E_o$ (equation \ref{Eo}) and $T_o$ (equation \ref{eq:To}), we are able to deduce a localization length $a\approx3nm$ and a density of states $g_o\approx10^{25}/eVm^3$ in our Na$_{2}$IrO$_{3}$ thin crystal devices. Localization length is about 10 times larger than the one reported for heteroepitaxial thin films, calculated using a density of states estimated for other iridates \cite{Jenderka}. Our analysis provides a value for both the localization length and the density of states in Na$_{2}$IrO$_{3}$ thin crystals deduced from the same set of data.

Mott's VRH requires a non-vanishing density of states at the Fermi level, as hopping between sites are concentrated in a band near the Fermi energy with a bandwidth $\epsilon_o$ (eqn. \ref{epsilon}). Our ARPES measurements confirm this in our samples.

\begin{figure}
\includegraphics[width=120mm]{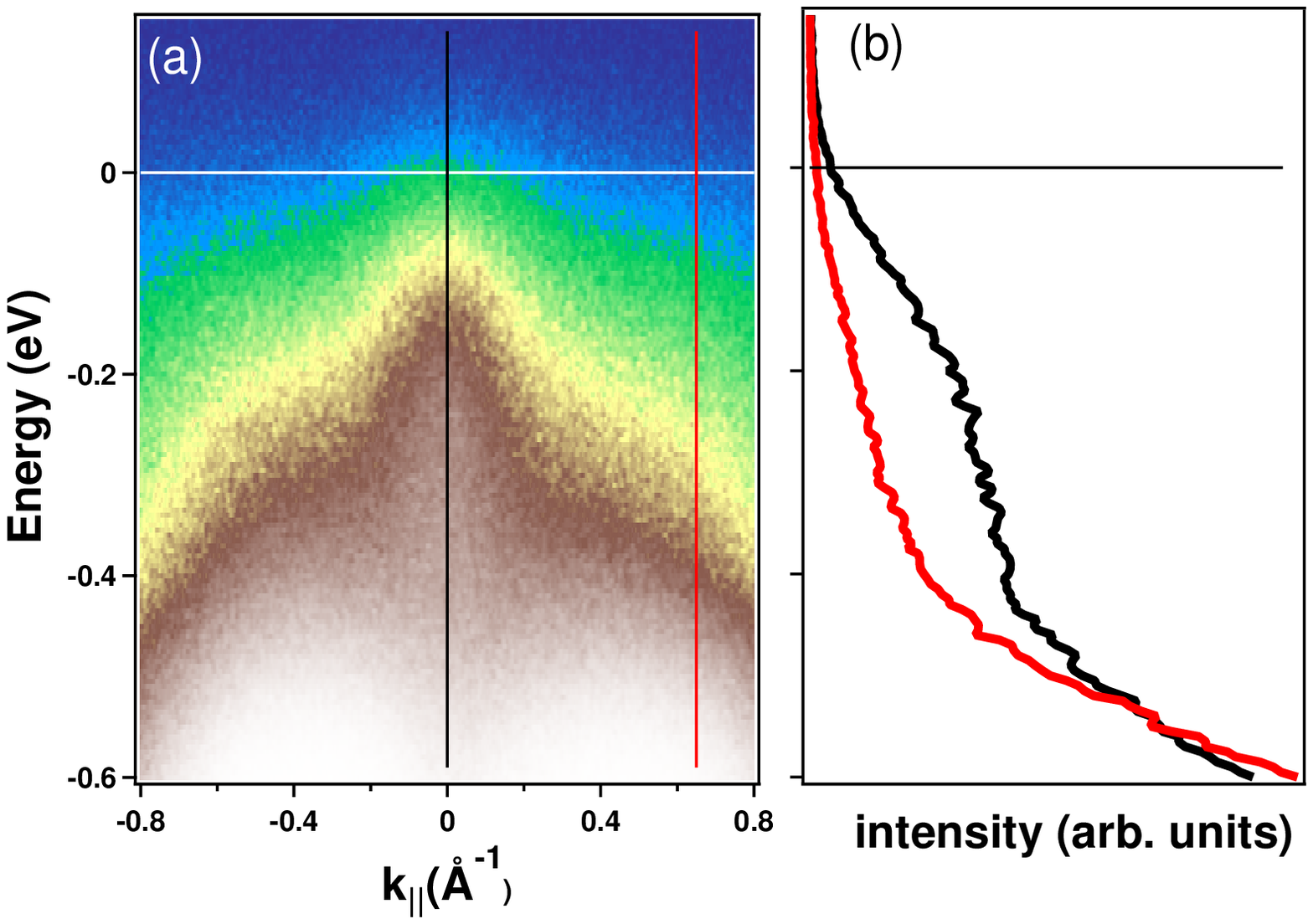} 
\caption{\label{fig:Fig5}\textbf{(a)} Measured electronic band structure along the high symmetry direction $\Gamma$-M with photon energy 90~eV  \textbf{(b)} Energy distribution curves integrated over 0.04\AA$^{-1}$ around $\Gamma$ (black) and at 0.65\AA$^{-1}$ as indicated in \textbf{(a)} by the red and black lines.} 
\end{figure}

ARPES measurements were performed at Beamline 4.0.3 (MERLIN) at the advanced light source using $90$ eV photons. Vacuum was better than $5\times 10^{-11}$ Torr. The total energy resolution was $20$ meV with angular resolution ($\Delta\theta\leqslant 0.2^{\circ}$). Data was taken at 290K to avoid charging of the sample. A fresh crystal of Na$_2$IrO$_3$ (as those used for the electronic device fabrication) was mounted and cleaved in-situ the ARPES chamber. Figure \ref{fig:Fig5}a shows the measured electronic band structure along the high symmetry direction $\Gamma$-M with a photon energy of 90eV. A feature near the Fermi energy is visible, as reported in previous ARPES works. Initially, this in-gap feature was identified as a metallic surface state \cite{Alidoust} but later on it was found through spatially resolved-ARPES measurements \cite{Moreschini} that it was consequence of quasiparticle formation in one of the possible cleavage planes of Na$_2$IrO$_3$. Indeed, the edge sharing IrO$_6$ octahedra in Na$_2$IrO$_3$ form a layered stacking alternating with pure Na layers, which gives rise to both Na and Ir-O terminated surfaces after cleavage. In the former, the Ir-O octahedra being embedded between two Na layers, create a charge transfer from Na to Ir giving rise to a quasiparticle at the Fermi level \cite{Moreschini}. This band formed at one of the possible terminations in Na$_2$IrO$_3$ after cleavage (the Na-terminated) is what we believe mediates the VRH mechanism in our Na$_2$IrO$_3$ devices. In the frame of Mott's VRH, the width of the band responsible for conduction (eqn. \ref{epsilon}) at 290K, corresponding to the ARPES measurements, is $\approx$ 90meV, which falls within the quasiparticle observed in Fig. \ref{fig:Fig5}. Fig. \ref{fig:Fig5}b shows energy distribution curves (EDC) in correspondance of the quasiparticle (in black, at $\Gamma$) and at a larger momentum outisde (in red) that testifies a non-vanishing density of states at the Fermi level, as required by Mott's VRH. 

After cleavage, the Na and O terminated surfaces are approximately 10-40 $\mu m$ large, as demonstrated through $\mu$ARPES \cite{Moreschini}. Our ARPES data being taken with a spotsize of $\approx 100\mu m$, captures an average of the two terminations. However, the spectral weight measured in the vicinity of the Fermi level is to be attributed to the Na-terminated surface only, since the Ir-O surface has no states in that energy range\cite{Moreschini}. This does not invalidate any of our conclusions and in fact supports them, since this density of states is also representative of the bulk, which we probe with our transport measurements. The band structure characteristic of the Ir-O cleave that lacks of states near the Fermi energy, is instead an "anomaly" given the absence of the Na layer which causes a difference charge balance between the two topmost layers of the cleave \cite{Moreschini}. Therefore, the electronic structure as measured by ARPES, both here and in refs. \onlinecite{Alidoust} and \onlinecite{Moreschini}, does show a small density of states within a small range from the Fermi level. 

In conclusion, we have integrated a thin crystal of Na$_2$IrO$_3$ into an electronic device, presenting the same Raman active modes as for the bulk crystals. We have observed that the main transport mechanism at temperatures above the formation of an antiferromagnetic long range order ($\approx$15K) is VRH. 
This observation is supported by ARPES measurements at the high temperature end (290K) that testify a non-vanishing density of states at the Fermi level as required by Mott's VRH. We have observed that as the system approaches long range order, it deviates from VRH with no evidence of this mechanism below the ordering temperature. Similarly, we have found that in the presence of an external electric field, transport diverges from VRH, becoming field assisted. Contrasting our data to the Mott's VRH and electric field assisted models allow us to deduce a localization length of $\approx 3nm$ and a density of states of $\approx10^{25}$/eVm$^3$ in our Na$_2$IrO$_3$ thin crystal devices. Our work constitutes a first approach to integrate an exfoliated thin crystal of Na$_2$IrO$_3$ into an electronic device, where separate ARPES measurements inform the electronic transport experiments. Using these two experimental techniques independently on similarly prepared samples can unveil important properties in other layered materials.

\begin{acknowledgments}
The primary funding for this work was provided by the U.S. Department of Energy, Office of Science, Office of Basic Energy Sciences under contract DE-SC0018154. The Advanced Light Source is supported by the Director, Office of Science, Office of Basic Energy Sciences, of the U.S. Department of Energy (U.S. DOE-BES) under contract no. DE-AC02-05CH11231. Work by J.G.A., G.L., V.N. and N.P.B. was supported by the Department of Energy \textit{Early Career Program}, Office of Basic Energy Sciences, Materials Sciences and Engineering Division, under Contract No. DE-AC02-05CH11231 for crystal growth. P. S and H. T. were supported by the National Institute of General Medical Sciences of the National Institutes of Health under Award Numbers UL1GM118979, TL4GM118980, and RL5GM118978 for Raman measurements. The content is solely the responsibility of the authors and does not necessarily represent the official views of the National Institutes of Health. We would like to acknowledge Luca Moreschini for valuable advice on the ARPES data as well as Jonathan Denlinger for assistance during the ARPES experiments.
\end{acknowledgments}


\end{document}